\documentclass[a4paper,11pt]{article}
\usepackage{pos}
\usepackage{epsfig}
\title{Gravitational waves from neutrino mass genesis}
\author*[a]{Pasquale Di Bari}
\affiliation[a]{Physics and Astronomy, University of Southampton,\\
  Southampton, SO17 1BJ, U.K.}

\emailAdd{P.Di-Bari@soton.ac.uk}
\abstract{The discovery of gravitational waves opens new opportunities to test BSM physics. 
In particular, the production of a stochastic background of primordial gravitational waves could
provide a signature of the generation of the right-right Majorana neutrino mass term 
necessary, within type-I seesaw mechanism, to explain lightness of neutrinos and their mixing parameters. 
I will discuss the possibility that such a generation occurs during a strong first order phase transition within Majoron models \cite{DiBari:2021dri}. 
As well known, this can indeed produce a stochastic
background of gravitational waves. The scale of the phase transition can or cannot coincide with the seesaw scale.
In the latter case a low scale phase transition, occurring in the pre-recombination era, 
might be tested at very low frequencies ($10^{-9}$--$10^{-6}\,{\rm Hz}$). Even though the signal can 
hardly reproduce the NANOGrac putative signal such new physics at low scale might
help ameliorating the tensions in the $\L$CDM cosmological model (e.g., the Hubble tension).
I will also discuss how a phase transition might be responsible for the generation of dark matter
in the form of dark neutrinos coupling to the seesaw neutrinos via Higgs induced right handed-right handed 
neutrino mixing \cite{DiBari:2020bvn}.}

\FullConference{%
  Corfu Summer Institute 2021 "School and Workshops on Elementary Particle Physics and Gravity"\\
  29 August - 9 October 2021\\
  Corfu, Greece
}

\begin{document}
\maketitle

\def\a{\alpha}
\def\b{\beta}
\def\c{\chi}
\def\d{\delta}
\def\e{\epsilon}
\def\f{\phi}
\def\g{\gamma}
\def\h{\eta}
\def\i{\iota}
\def\j{\psi}
\def\k{\kappa}
\def\la{\lambda}
\def\m{\mu}
\def\n{\nu}
\def\o{\omega}
\def\p{\pi}
\def\q{\theta}
\def\r{\rho}
\def\s{\sigma}
\def\t{\tau}
\def\u{\upsilon}
\def\x{\xi}
\def\z{\zeta}
\def\D{\Delta}
\def\F{\Phi}
\def\G{\Gamma}
\def\J{\Psi}
\def\L{\Lambda}
\def\O{\Omega}
\def\P{\Pi}
\def\Q{\Theta}
\def\S{\Sigma}
\def\U{\Upsilon}
\def\X{\Xi}

%Varletters
\def\ve{\varepsilon}
\def\vf{\varphi}
\def\vr{\varrho}
\def\vs{\varsigma}
\def\vq{\vartheta}

\def\dg{\dagger}                                     % hermitian conjugate
\def\ddg{\ddagger}                                   % double dagger
\def\wt#1{\widetilde{#1}}                    % big tilde
\def\mt{\widetilde{m}_1}
\def\mti{\widetilde{m}_i}
\def\rt{\widetilde{r}_1}
\def\mtt{\widetilde{m}_2}
\def\mttt{\widetilde{m}_3}
\def\rtt{\widetilde{r}_2}
\def\mb{\overline{m}}
\def\VEV#1{\left\langle #1\right\rangle}        % < >
\def\be{\begin{equation}}
\def\ee{\end{equation}}
\def\ds{\displaystyle}
\def\ra{\rightarrow}

\def\bea{\begin{eqnarray}}
\def\eea{\end{eqnarray}}
\def\NO{\nonumber}
\def\Bar#1{\overline{#1}}

\section{Introduction}

The expectations of discovering new physics at the TeV scale in colliders have so far been disappointed.
Yet, we know from the discovery of neutrino masses and mixing and from cosmological observations that
current established description of fundamental laws of physics, 
based on general relativity and standard model of particle physics, 
needs to be extended to some level. The kind of new physics able to address the cosmological puzzles 
and explain neutrino masses then must be either  at energy scales higher than
those accessible to colliders or hide in a way not to give detectable effects, for example residing
in a very weakly coupled sector, or some combination of the two. 

Fortunately, novel important phenomenological tools open new opportunities to explore models of new physics 
at energy scales much higher than the TeV scale and/or containing very weakly coupled sectors. 
Gravitational waves (GW) certainly represent a very promising tool in this respect.  
The production of different kinds of stochastic primordial backgrounds of gravitational waves can indeed occur in the early universe within many different scenarios of new physics. This is 
a list of well known sources of GW stochastic primordial backgrounds associated to new physics:
\begin{itemize}
\item Vibration of cosmic strings and domain walls \cite{Vachaspati:1984gt};
\item During inflation \cite{Grishchuk:1974ny,Starobinsky:1979ty};
\item At preheating \cite{Khlebnikov:1997di};
\item From the dynamics of extra-dimensions \cite{Hogan:2000is};
\item From primordial black holes \cite{Mandic:2016lcn};
\item From Affleck-Dine \cite{White:2021hwi};
\item From strongly first order phase transitions (SFOPT) \cite{Witten:1984rs,Hogan:1986qda,Turner:1990rc}.
\end{itemize}
In my talk I will focus on the last mechanism. As I will discuss, SFOPTs can quite 
naturally be associated to the origin of neutrino masses. 

\section{GW from SFOPTs as a signature of new physics}

The production of stochastic background of primordial GWs from SFOPTs were first studied within the SM,  both
from a phase transition associated to chiral symmetry breaking \cite{Witten:1984rs} 
and from an electroweak symmetry breaking (EWSB) phase transition \cite{Kamionkowski:1993fg}. 
However, we know today that in  both cases  symmetry breaking would occur as a smooth crossover in the SM.
For this reason the detection of primordial GWs would be then provide a signature of new physics. 

Most of the attention has then focused on EWSB in extensions of the SM, especially in connection
with electroweak baryogenesis and mainly within supersymmetric models: in the MSSM \cite{Apreda:2001us}, 
in the NMSSM \cite{Pietroni:1992in}, in generic extensions of the SM with gauge singlets \cite{Choi:1993cv}.
Of course this was  in addition also presenting
the possibility to incorporate the WIMP miracle in a way to provide a very attractive package able to 
address the problem of the origin of matter  (dark matter and matter-antimatter
asymmetry) in the universe \cite{origin} and, on more theoretical grounds, the naturalness problem. 

However, with the (Run 1 + Run 2) LHC constraints on BSM physics at the electroweak scale,
a SFOPT associated to EWSB, though not completely excluded, does not seem as compelling as in the
pre-LHC era and we are now in a kind of {\em nothing is impossible} or agnostic time  
concerning the scale (or scales)  of new physics, being driven mainly by experimental anomalies and hints rather 
than theoretical arguments.   This will also be the approach I will follow in this talk. 
 
\section{Scale of new physics from GW stochastic backgrounds} 
 
 Let us introduce the GW spectrum contribution to the energy density parameter    
\be
h^2\,\O_{GW 0}(f)=  {1 \over \rho_{{\rm c} 0}h^{-2}} \,  {d\rho_{GW 0}\over d\ln f} \,  . 
 \ee 
This will typically contain some characteristic frequency $\bar{f}_\star$ at the time of the phase transition $t_\star$ that
in a standard cosmology is redshifted at the present time $t_0$ at the frequency
\be
\bar{f}_0 = \left({g_{S0}\over g_{S\star}}\right)^{1/3}\,{T_0 \over T_\star} \, \bar{f}_\star
 \simeq 6\times 10^{-3}\,{\rm mHz} \, \frac{\bar{f}_\star}{H_\star}  \frac{T_\star}{\rm 100 GeV} 
 \left(\frac{g_\star}{106.75} \right)^{1/6} \,  .
\ee 
In this way one can probe the scale of new physics $T_\star$ that can be extracted from the signal 
knowing $\bar{f}_\star/H_\star$, where $H_\star$ is the expansion rate at $t_\star$. The numerical
expression highlights that  for transitions at the electroweak scale $T_\star \sim 100\,{\rm GeV}$
and for typical values $f_\star/H_\star \sim 100$ one expects a signal in the mHz range that will be
tested by experiments such as LISA \cite{Caprini:2015zlo}. 

In the specific case of GWs from SFOPTs, the characteristic frequency can be easily identified  with the
{\em peak frequency} $f_{\rm peak}$.  This provides a very clear way to extract the scale of the phase
transition $T_\star$, though there are some parameters entering the calculation that need to be known for such connection.
In general, there are three contributions to $h^2\,\O_{{\rm GW},0}(f)$:
from bubble wall collisions, from sound waves and from turbolence \cite{Caprini:2015zlo}.  
In Fig.~1 one can see a few examples showing how higher peak frequencies correspond to higher scales $T_\star$ and
calculated assuming dominance of sound wave contribution. 
\begin{figure}
\begin{center}
        \psfig{file=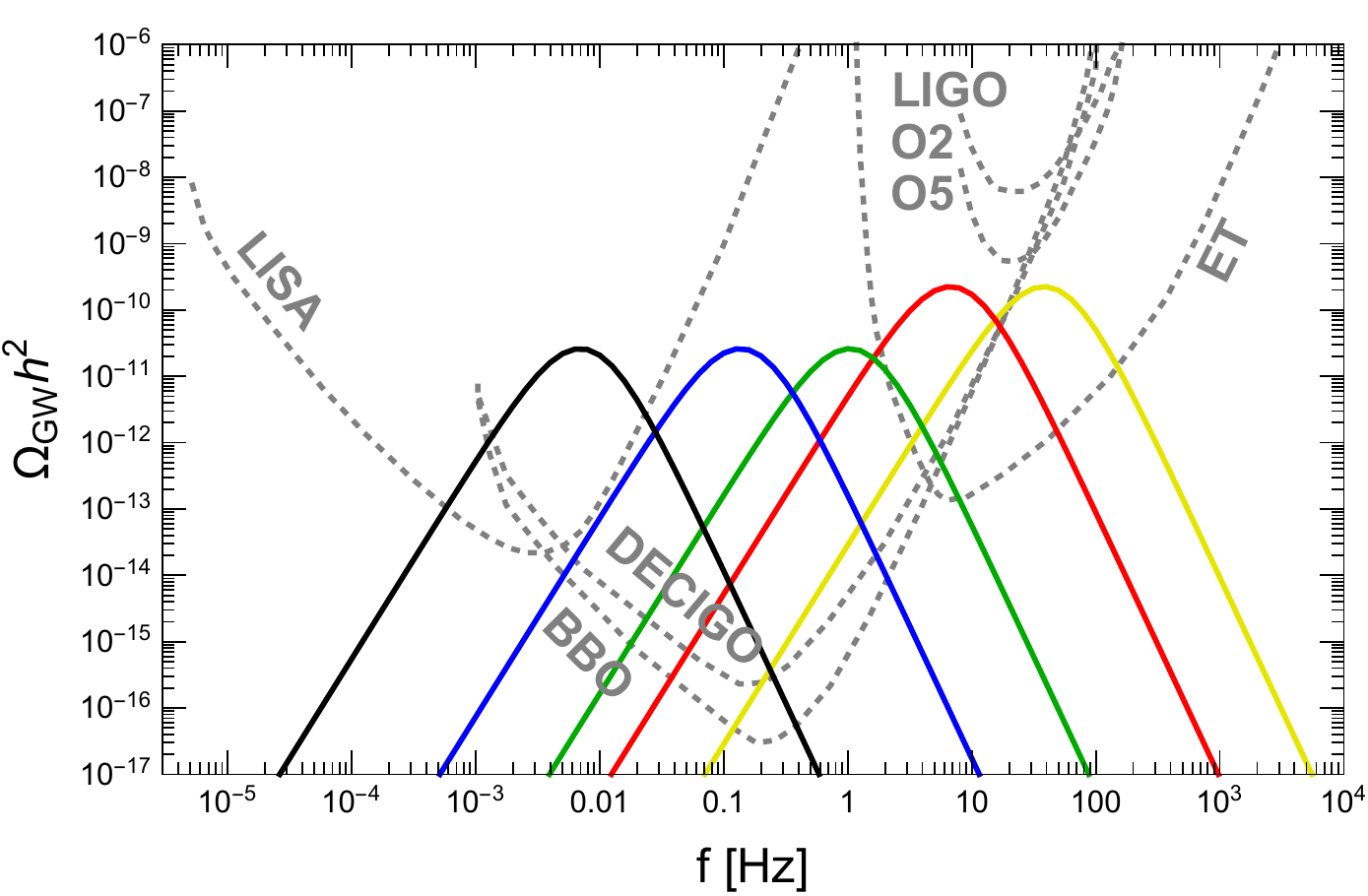,height=65mm,width=89mm}
        %\caption{resonant case}
        %\label{fig:gws}
\caption{Examples of GW spectra dominated by the sound wave contribution (from \cite{DiBari:2020bvn}).}
\label{fig:gws}
\end{center}
\end{figure}
I will now discuss how such a spectrum can be obtained from a model of first order phase transition. 

\section{First order phase transitions in the early universe}

I now briefly review the main points in the description of a phase transition in the early universe. 
Given a scalar field $\phi$, there will be an associated tree level zero temperature potential $V_{\rm tree}(\phi)$.
This describes the broken symmetry phase of the particle sector undergoing the phase transition that in our case, as we will
discuss, it will be a dark sector.
At high temperatures one has to include thermal effects that are usually described by one-loop effective effective
potential at finite temperatures $V_1^T(\phi)$. This includes both a zero temperature one-loop contribution, 
given by the Coleman-Weinberg potential $V_{\rm CW}(\phi)$, and the one-loop thermal potential $V_T(\phi)$.
In this way one obtains the finite-temperature effective potential at one-loop as a sum of three contributions
\cite{Kirzhnits:1972ut,Dolan:1973qd}:
\be
V_{\rm eff}^T(\phi) = V_{\rm tree} (\phi) + V_{\rm CW}(\phi) + V_T(\phi) \,  .
\ee
Above a certain critical temperature $T_{\rm c}$, defined as that temperature where the initial high temperature 
minimum at zero  field value becomes degenerate with the symmetry breaking minimum at nonzero field value,  
one typically has symmetry restoration. 

In electroweak baryogenesis a high-temperature expansion leads to a polynomial form for the
finite-temperature effective potential at one-loop \cite{Anderson:1991zb,Dine:1992wr,Quiros:1999jp}
\be\label{VTeffminimalEWB}
 V^T_{\rm eff}(\s_1) \simeq D\, (T^2 - T_0^2) \s_1^2 - A \, T \, \s_1^3 + \frac{1}{4}\lambda_T\, \s_1^4 \,  .
 \ee
The probability of bubble nucleation per unit volume and time at temperature $T$ can be written in the form
\cite{Coleman:1977py,Linde:1981zj}
\be
\Gamma(T) = \bar{\Gamma}(T)\,e^{-S_E(\phi,T)} \,  ,
\ee
where $\bar{\Gamma}(T) = {\cal O}(1)\,T^4$ and $S_E(T)$ is the Euclidean action
\be
S_E(\phi,T) = \int d\tau d^3 x \, \left[{1\over 2}\left({d\phi \over d\tau}\right)^2 
+ {1\over 2}\,\nabla^2 \phi + V_{\rm eff}^T(\phi) \right] \,  .
\ee
The Euclidean action has to go to infinity for $T \ra T^-_{\rm c}$ since in this way there is no 
bubble nucleation above the critical temperature.  For temperatures higher than the
inverse bubble radius at zero temperature $r_{0}^{-1}$ \cite{Linde:1981zj}, one has
\be
\lim_{T \gg r_{0}^{-1}} S_E(\phi,T) = {S_3(\phi,T) \over T} \,  ,
\ee
where $S_3(\phi,T)$ is the spatial Euclidean action given by
\be
S_3(\phi,T) = 4\pi\,\int dr \, r^2 \, \left[ {1\over 2}\left({d\phi \over dr}\right)^2 +  V_{\rm eff}^T(\phi) \right] \,  .
\ee
The Euler-Lagrange equation is then simply 
\be
{1\over 2}\left({d\phi \over dr}\right)^2 + {2\over r}\,{d\phi \over dr} - 
{\partial V_{\rm eff}^T(\phi) \over \partial T} = 0 \,  .
\ee
Imposing boundary conditions $\phi(r=\infty)=0$  and $(d\phi/dr)_{r=0} = 0$, 
one finds numerical solutions by overshooting-undershooting trials and errors procedure 
in the form of bounce solutions. In the {\em thin-wall approximation}, a kink solution is found  analytically:
\be
\phi(r,t) = {1\over 2}\,\langle \, \phi \, \rangle \, 
\left[ 1 -\tanh\left({r-r_{\rm n}-v_{\rm w}(t-t_{\rm n})\over \D_{\rm w}}\right)\right] \,  ,
\ee
where $v_{\rm w}$ and $\D_{\rm w}$ are respectively the bubble wall velocity and thickness and
$t_{\rm n}$ is the nucleation time of the bubble.  The nucleation time is defined as that time 
such that 
\be
\int_0^{\rm t_{\rm n}}\,dt \,{\Gamma \over H^3} = 1 \, ,
\ee
and from this one arrives to the condition \cite{Grojean:2006bp}
\be
{S_3(T_{\rm n})\over T_{\rm n}} = -4\ln\left({T_{\rm n}\over M_{\rm P}}\right) \,  ,
\ee
that gives the temperature at the nucleation time $T_{\rm n}$.  One can also define
a {\em percolation time} $t_\star$ that can be identified with the time of the phase transition
where to calculate the parameters entering the calculation of the GW spectrum. It is defined
as the time when the false vacuum has filled a fraction $1/e$ of space \cite{Megevand:2016lpr}. In any case 
we are interested in phase transitions
where the duration is very short compared to the life of the universe. In this case  an exact definition of
the time $t_\star$ and temperature $T_\star$  of the phase transition where to calculate the parameters
that enter the calculation of the GW spectrum gives small differences in the calculation of the GW spectrum.
If we define $\beta \equiv \dot{\Gamma}/\Gamma$, then this can be calculated around $t_\star$, finding
$\beta \simeq -(dS_E/dt)_{t_\star}$ and from this one finds an expression to calculate the important
parameter $\beta/H_\star$ from the Euclidean action
\be
{\beta\over H_\star} \simeq T_\star \, \left.{d(S_3/T)\over dT}\right|_{T_\star} \,  .
\ee
The other important parameter is the {\em strength of the phase transition}
\be
\a \equiv {\ve(T_\star) \over \rho_{R}(T_\star)} \,  ,
\ee
where $\ve(T_\star)$ is the latent heat freed in the phase transition and $\rho_{R}(T_\star)$
is the radiation energy density of the plasma at $t_\star$.
Having $\a$  and $\beta/H_\star$, one can calculate the GW spectrum within certain approximations.  

\section{GW spectrum from FOPTs}

As we anticipated, there are three established contributions to the GW spectrum \cite{Caprini:2015zlo}.
Let us focus on the sound wave contribution, since in our case this proves to be the dominant one.  
We can then write
\be\label{omegasw}
h^2\Omega_{\rm sw}(f) =2.59 \times 10^{-6} \, \frac{v_{\rm w}(\a)}{\beta/H_\star} \left[\frac{\kappa(\a)\, \alpha}{1+\alpha}\right]^2  
\,\left( \frac{106.75}{ g_\rho^\star} \right)^{1/3} S_{\rm sw} (f) \,,
\ee
where $g_\rho^\star \equiv g_\rho(T_\star)$ is the number of (energy density) ultrarelativistic degrees of freedom at the phase transition. The quantity $S_{\rm sw} (f) $ is
the {\em spectral shape function} and is given by
\begin{eqnarray}
S_{\rm sw} (f) = \left(\frac{f}{f_{\rm sw}}\right)^3 \left[\frac{7}{4+3({f/f_{\rm sw}})^2} \right]^{7/2} \,  ,
\end{eqnarray} 
with the peak frequency 
\begin{eqnarray} \label{fpeak}
f_{\rm peak} = f_{\rm sw} \simeq 1.92\times 10^{-2}\,{\rm mHz} \, \frac{1}{v_{\rm w}(\a)} \frac{\beta}{H_\star}  \frac{T_\star}{\rm 100\,GeV} \left( \frac{g_\rho^\star}{106.75} \right)^{1/6} \, .
\end{eqnarray}
In the expression (\ref{omegasw}) the quantities $\kappa(\a)$
and $v_{\rm w}(\a)$ are the efficiency factor and the bubble wall velocity respectively.
The efficiency factor $\kappa(\a)$ measures how much of the vacuum energy is converted to bulk kinetic energy. 
Adopting Jouguet detonation solutions, the efficiency factor can be calculated as~\cite{Steinhardt:1981ct} 
\begin{eqnarray}
\kappa(\a) \simeq \frac{\sqrt{\alpha}}{0.135+\sqrt{0.98+\alpha}} \,,
\label{efficiencyfactor}
\end{eqnarray}
and the bubble wall velocity as $v_{\rm w}(\a) = v_{\rm J}(\a)$,  where
\begin{eqnarray} \label{eq:Jouguet}
v_{\rm J}(\a) \equiv \frac{\sqrt{1/3} + \sqrt{\alpha^2 +2\alpha/3}}{1+\alpha}\,.
\end{eqnarray}
Jouguet solutions provide a simple and useful prescription. Within a more rigorous treatment, the bubble velocity deviates from $v_{\rm J}(\a)$, and the efficiency factor is a function of both $\a$ and $v_{\rm w}$. Moreover, the friction exerted by the plasma on the bubble wall also needs to be taken into account, leading to a much more complicated description that requires numerical solutions of the Boltzmann equations~\cite{Espinosa:2010hh}.  It should be said that
this is just one of many sources of theoretical uncertainties in the calculation of the GW spectrum.  We will comment more
on this important point in the summary.

\section{First order phase transition associated to Majorana mass generation in the Majoron model}

The SM needs to be extended to incorporate neutrino masses and mixing. In a minimal extension of the SM,  one
adds a Dirac mass term as for the other massive fermions. This of course implies that RH neutrinos
need to be included in the particle content, though these would be gauge singlets and, therefore, behave as sterile neutrinos. 
However, this minimal extension does not address the lightness of neutrinos compared to the other massive fermions
and it cannot address also the problem of the origin of matter in the universe. Moreover, unless, one imposes
lepton number conservation as a symmetry of the theory, a right-right Majorana mass term should also be added. 
In this way, after spontaneous symmetry breaking, one has the following neutrino mass term in the Lagrangian: 
\be\label{seesawL}
-{\cal L}^{\nu}_{\rm m}= {1\over 2}\,
\left[(\overline{\nu_L},\overline{\nu_R^{c}})
\left(
\begin{array}{cc}
      %\bo{1}
                0  & m_D  \\
               m_D^T &  M    \\
\end{array}\right)
\left(
\begin{array}{c}
       \nu_L^{c}  \\
       \nu_R  \\
\end{array}\right)
\right] + {\rm h.c.} \,  .
\ee
Since this is in the form of a Majorana mass term, all physical fields are Majorana fields. In the seesaw limit,
for $M\gg m_D$, the mass spectrum splits into a heavy set with masses approximately coinciding with
the eigenvalues of $M$ and a light set corresponding to ordinary neutrinos, with a light neutrino mass
matrix given by the seesaw formula 
 \be\label{seesawdiagonal}
D_m =  U^{\dagger}\,m_D \, {1\over D_M} \,m_D^{T} \, U^{\star} \,  ,
\ee
written here in the flavour basis where both charged lepton and Majorana mass matrices are diagonal
and where $D_m \equiv {\rm diag}(m_1,m_2,m_3)$ are light neutrino masses and 
$D_M \equiv {\rm diag}(M_1,M_2,\dots,M_N)$ are the heavy neutrino masses. Notice that the
number of RH neutrinos $N \geq 2$ for the seesaw formula to successfully reproduce low energy neutrino data.

The nice thing is that such an extension can  both explain the lightness of neutrinos and address the problem of the origin 
of matter in the universe with leptogenesis and with one sterile neutrino as dark matter. 
The Majorana mass term is included into the Lagrangian (\ref{seesawL}) but one could wonder whether this could also
originate from spontaneous symmetry breaking and in this case a first order phase transition with a consequent
GW production becomes a natural possibility. 

This can be done within a Majoron model \cite{Chikashige:1980ui}. 
In this case RH neutrinos couple not only to lepton doublets
with Yukawa couplings but also to a complex scalar singlet
\be
\sigma ={1 \over \sqrt{2}}\,(\s_1 + i \, \s_2) \,  ,
\ee
with couplings $\lambda_I$, so that one has
\bea\label{eq:L_l}
-{\cal L}_ {N_I+\s} & = & 
 \overline{L_{\a}}\,h_{\a I}\, N_{I}\, \widetilde{\Phi} 
+  {\lambda_{I}\over 2}  \, \sigma \, \overline{N_{I}^c} \, N_{I}
+ V_0(\sigma)
+ {\rm h.c.} \, .
\eea
The potential $V_0(\sigma)$ drives the phase transition of $\sigma$. During the phase transition, for $T \sim T_\star$,
the vev is given by $\langle \sigma \rangle = v_T/\sqrt{2}$. After SSB the field will evolve into
\be\label{sigma}
\sigma ={e^{i\theta} \over \sqrt{2}}\,(v_0 +  S + i \, J) \,  ,
\ee 
and Majorana masses $M_I = \lambda_I\,v_0/\sqrt{2}$ are generated. Here $S$ is a 
massive  field with $m_S^2= 2 \lambda v_0^2$ 
and $J$ is  the {\em Majoron}, a massless Goldstone field. In what follows
we will talk of a {\em seesaw scale} $M$, that corresponds either to a common mass
of the RH neutrinos, assumed to be quasi-degenerate, or to the mass of the heaviest, since this dominates in 
the effective thermal potential. 

At the moment let us assume $T_\star > v_{\rm ew}$, so that the $\sigma$-phase transition occurs
prior to the electroweak phase transition. 
After electroweak symmetry breaking a Dirac neutrino mass matrix $m_D = v_{\rm ew}h/\sqrt{2}$
will also be generated and again the light neutrino mass matrix will be given by the seesaw formula. 

We have now to calculate the thermal effective potential of the model. Given the measured values of solar and atmospheric neutrino mass scales, RH neutrinos thermalise prior to the $\sigma$-phase transition and contribute to 
$V_{\rm eff}^T(\sigma)$.  In what follows we will refer to the {\em dark sector}
as the set of RH neutrinos plus $S$ and $J$ and to the {\em visible sector} as the set of
SM particles. 
 
\section{The minimal model}

Let us first start considering a minimal model where $(\mu^2,\lambda > 0)$
\begin{eqnarray} \label{eq:V_L}
V_0(\sigma) = - \mu^2 |\sigma|^2 + \lambda |\sigma|^4    \;\;\; \,  .
\end{eqnarray}
The zero temperature vev of $\sigma$ is in this case given by $v_0 = \sqrt{\mu^2/\lambda}$ while the
mass of $S$ is easily found to be $m^2_{\rm S} = 2\lambda\,v_0$. Notice that interactions
$N_I + N_I \leftrightarrow \sigma$ 	thermalise $\sigma$, and therefore $S$ and $J$, 
prior to the phase transition. One has then to arrange $m_S$ in a way not to be lighter than all
RH neutrinos so that it can decay into one of them, otherwise its abundance would overclose the universe.

Similarly to the case of electroweak baryogenesis, a high temperature expansion of the thermal functions in the expression of
the one-loop finite temperature effective potential yields an approximate polynomial form:
\be\label{VTeffminimal}
 V^T_{\rm eff}(\s_1) \simeq D\, (T^2 - T_0^2) \s_1^2 - A \, T \, \s_1^3 + \frac{1}{4}\lambda_T\, \s_1^4 \,  ,
 \ee
 where notice that, without loss of generality, one can always 
 study the problem along the real axis so that $\sigma \ra \sigma_1$.
In this expression the {\em destabilisation temperature} $T_0$, marking the
end of the phase transition, is given  by %%
\be
2\,D\,T_0^2 =   \lambda\,v_0^2 +{N \over 8\,\pi^2}\,{M^4 \over v_0^2} 
-{3\over 8 \,\pi^2}\lambda^2 \, v_0^2  \,  ,
\ee
where the dimensionless coefficients $D$ and $A$ are
\be\label{DA}
D = {{\lambda \over 8} + {N\over 24}\,{M^2 \over v_0^2}} \,   \;\;\;\; \mbox{\rm and} \;\;\;\;
A =  {(3\,\lambda)^{3/2} \over 12\pi } \,  \,  .
\ee
 The cubic term  is obtained using the approximation, 
$m^2_{\s}(\s_1) = \Pi_\s -\lambda v_0^2+3\lambda \s_1^2 \simeq 3\lambda \s_1^2$. This approximation
works quite well since the cubic term gives a nonnegligible contribution 
only when the field is large. Finally, the dimensionless temperature dependent 
coefficient $\lambda_T$ is given by
\be\label{lambdaT}
\lambda_T = \lambda  - \frac{N\, M^4}{8\,\pi^2 \, v_0^4}  \, \log {a_F \, T^2 \over e^{3/2}\,M^2} 
+  {9\lambda^2 \over 16 \pi^2}\,\log {a_B \, T^2 \over e^{3/2}\,m_S^2} \,  .
\ee
Here it is interesting to notice that stabilisation requires $\lambda_T > 0$
and this places an upper bound on the number of RH neutrinos $N$.

It is then possible to derive an expression for the Euclidean action in the form 
\be\label{euclidean}
{S_3 \over T} = {\widetilde{M}_T^3 \over A^2 \, T^3} \, f(a) \,  ,
\ee
where we introduced
\be
\widetilde{M}_T^2\equiv 2\,D\,(T^2 - T_0^2) \,   \;\;\; \mbox{\rm and} \;\;\;
a \equiv {\lambda_T \, \widetilde{M}_T^2 \over 2\,A^2 \, T^2 } \,  ,
\ee
and where the function \cite{Dine:1992wr}
\be
f(a)  \simeq 4.85\,\left[1 +{a \over 4}\,\left(1 +{2.4 \over 1-a} + {0.26 \over (1-a)^2} \right)\right] \,   
\ee
provides an accurate analytical fit. 

In the minimal model there are three parameters, $v_0$,  $m_{\rm S}/v_0$,  $M/v_0$. 
The results for  $\a$ and $\beta/H_\star$ of a scatter plot show that $\a \lesssim 10^{-3}$
and $\beta/H_\star \gtrsim 10^6$ and this  implies that
the peaks of the GW spectra one obtains are six or seven orders of magnitude below the sensitivity
of any planned experiment. Therefore, the minimal model cannot produce any detectable signal.

\section{Adding an auxiliary scalar} 

It is well known in the study of electroweak baryogenesis \cite{Choi:1993cv,Kehayias:2009tn} that the addition
of an auxiliary scalar can significantly enhance the strength of the phase transition. 
Introducing a very heavy real scalar $\eta$, the tree level potential can be written 
in general as
\be
V_0(\sigma,\eta) = V_0(\sigma) + V_{\eta\sigma}(\eta,\sigma) + V_{\eta}(\eta) \,  .
\ee
The most important term is contained in $V_{\eta\sigma}$, that can be written as
\be
V_{\eta\sigma} = {\delta_1 \over 2}\, |\sigma|^2 \, \eta + {\delta_2 \over 2}\, |\sigma|^2 \, \eta^2 \,  .
\ee
There are many different possibilities how symmetry breaking can occur when there is an interplay between two fields.
However, a simple (and  well be motivated as I will comment later on) case is that $\eta$ undergoes
its own phase transition and settles to its true vacuum prior to the $\sigma$-phase transition. In this case
a zero temperature barrier between the false and true vacuum appears and 
the one-loop effective potential of the minimal model in Eq.~(\ref{VTeffminimal}) gets modified into
\be\label{VTeffminimal}
 V^T_{\rm eff}(\s_1) \simeq D\, (T^2 - T_0^2) \s_1^2 - (A \, T +\widetilde{\mu}) \, \s_1^3 + 
 \frac{1}{4}\lambda_T\, \s_1^4 \,  ,
 \ee
where the coefficient in the new cubic term $\widetilde{\mu} \propto \delta_2$.        

The expression (\ref{euclidean})  for the Euclidean action in the minimal model now gets 
generalised with the replacement
\be
A^2 \, T^2  \rightarrow  \widetilde{\mu}_T \equiv A\,T + \widetilde{\mu} \,  ,
\ee
so that one obtains
\be\label{euclidean}
{S_3 \over T} = {\widetilde{M}_T^3 \over  \widetilde{\mu}_T^2 \, T} \, f(\widetilde{a}) \,  , \;\;
\mbox{\rm with} \;\;
\widetilde{a} \equiv   {\lambda_T \, \widetilde{M}_T^2 \over 2\,\widetilde{\mu}_T^2} \,  .
\ee
The presence of a non-vanishing zero temperature cubic term greatly enhances the
strength of the phase transition and accordingly the GW signal (as in the case of the
electroweak phase transition).

This time the results of a scatter plot show that there are particular choices of the
parameters for which the GW spectrum is within the sensitivity of planned experiments.
In Fig.~2 three of such spectra (blu curves) are shown.
\begin{figure}
\begin{center}
        \psfig{file=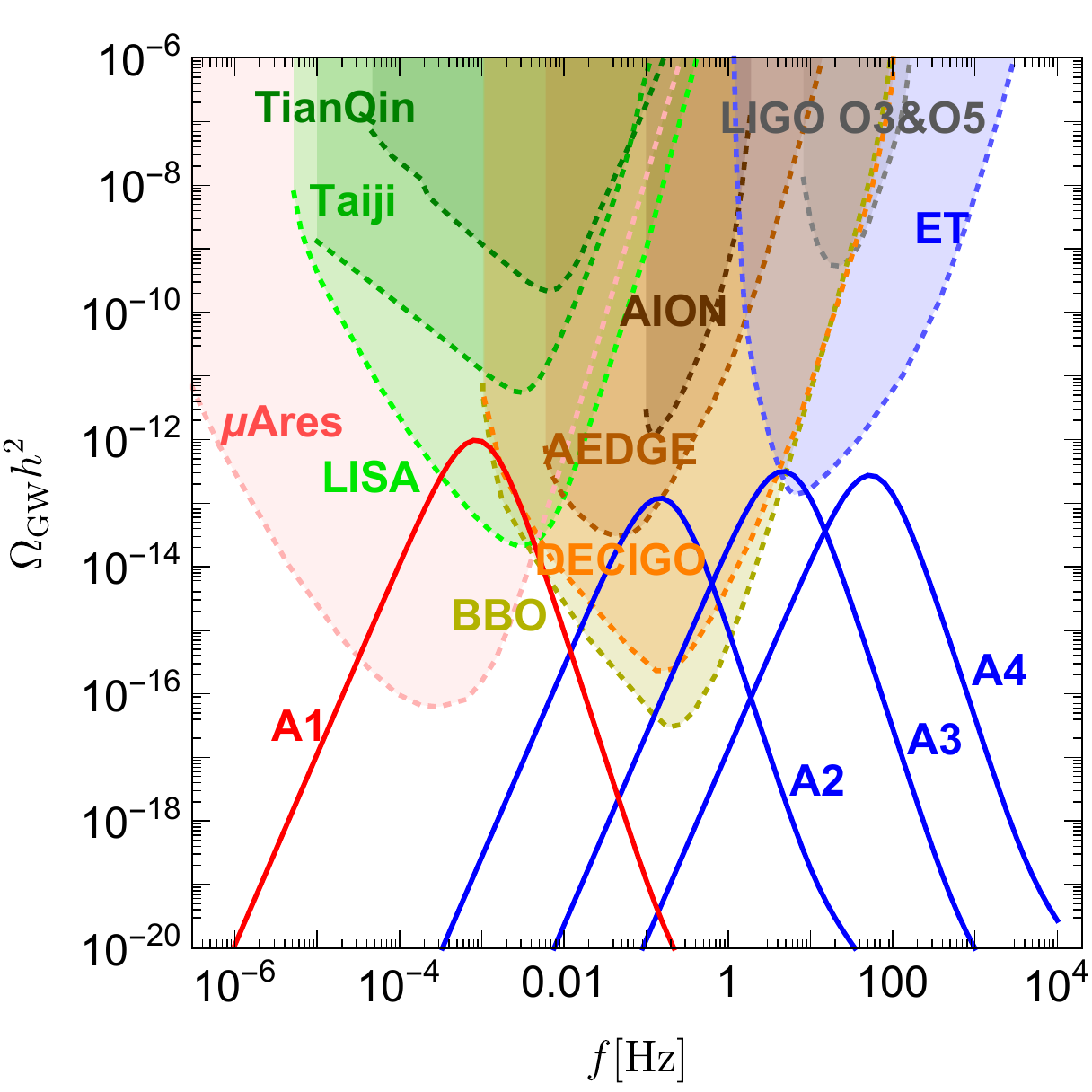,height=65mm,width=89mm}
\caption{GW spectra for the four benchmark points in the table. The blue curves correspond to
the case $T_\star > v_{\rm ew}$ while the red point correspond to a GeV seesaw scale scenario (from \cite{DiBari:2021dri}).}
\label{fig:gws}
\end{center}
\end{figure}
These correspond to the points A2, A3, A4 for values of the parameters shown in the Table. 
\begin{table}[t]
\begin{center}
%\begin{tabular}{ llllllll }
%\begin{table}[h]
%\begin{center}
\begin{tabular}{ l | cccc | cccc }
 \hline\hline
& \multicolumn{4}{c|}{Inputs} & \multicolumn{4}{c}{Predictions} \\
 \cline{2-9}
 &  $m_S/{\rm GeV}$ & $\tilde{\mu}/{\rm GeV}$ & $M/{\rm GeV}$ & $v_0/{\rm GeV}$ & $T_\star/{\rm GeV}$ & $\alpha$ & $\beta / H_\star$ & $\widetilde{a}$  \\\hline\hline
 A1 & $0.06190$ & $0.0005857$ & $0.5361$ & $3.5873$ & $0.6504$ & $0.1248$ & $2966$ & $0.05951$ \\
 \hline
A2 & $156.2$ & $13.15$ & $465.6$ & $1014$ & $721$ & $0.04139$ & $754.8$ & $0.3886$ \\
A3 & $1036$ & $13.72$ & $7977$ & $44424$ & $9180$ & $0.08012$ & $1975$ & $0.06268$ \\
A4 & $43874$ & $1856$ & $181099$ & $567378$ & $247807$ & $0.05611$ & $809.7$ & $0.1944$ \\
 \hline\hline
\end{tabular}
\end{center}
\caption{Values of the parameters corresponding to the four benchmark (starred) points in Fig.~\ref{f2}.}
%The first four correspond to the stars in Fig.~\ref{fig:scatterplot}, while the fifth is off scale and is not shown in Fig.~\ref{fig:scatterplot}.
\label{tab:samples_highT1}
\end{table}
The point A1 corresponds to the GW spectrum in red in Fig.~2 but this corresponds to the 
case low scale seesaw scenarios, with $T_\star$ below the electroweak scale,
that I am now going to discuss.

\section{Low scale scenarios}

We have so far assumed:
\begin{itemize}
\item[(i)] $T_{\star} \gtrsim v_{\rm ew}$;
\item[(ii)] $T_{\star} \sim M$.
\end{itemize}
Let us first relax (i)  considering the case $T_\star \sim M \ll v_{\rm ew}$.
In this case the dark sector can remain coupled to the visible sector
to a common temperature $T$ only if $M \gtrsim 1 \, {\rm GeV}$. 

\subsection{GeV seesaw scale seesaw scenarios}

Let us start considering the borderline case $M \sim {\rm GeV}$.
The only modification is that this time $g_\rho^\star \simeq 60$. However this increases
the value of $\a$ that approximately doubles. One can see clearly from Eq.~(\ref{omegasw})
that the GW spectrum increases at least as $\a^3$ when $\a$ increases. In fact the depensence is even stronger, since the parameter $\beta/H_\star$ is not independent of $\a$ and its minimum is approximately
$\propto \a^{-2}$ \cite{Ellis:2020awk}. Consequently, the peak of the GW spectrum increases approximately as $\a^5$.
This explains why the spectrum in Fig.~2 (red curve),
corresponding to a GeV seesaw scale scenario and 
to the point A1 in the table, has a peak that is about 30 times higher than the 
GW spectra for the high scale scenarios. 

It can be seen that the peak lies at mHz frequencies, well within LISA sensitivity. From this point of view notice that usually one thinks of  electroweak phase transitions, where $T_\star \sim 100\,{\rm GeV}$, 
falling in this frequency range. However, in that case the value of $\beta/H_\star \sim 10$ while in our case 
 $\beta/H_\star \sim 1000$ and from Eq.~(\ref{fpeak}) one can see that there is a shift toward higher frequencies for fixed $T_\star$. We will come back on this point.  

\subsection{Splitting the seesaw and phase transitions scales}

So far we considered the case that the scale of phase transition and the seesaw scale coincide. Let us 
now  discuss the possibility that the phase transition scale is lower than the seesaw scale, i.e.,
$T_\star < M$. We will still impose $M > {\rm GeV}$. This is a reasonable condition to impose 
since in this way some scenario of leptogenesis can reproduce the matter-antimatter asymmetry.  
The dark sector participating to the phase transition is now given by $N' = N-N_{\rm seesaw}$ light RH neutrinos while the visible sector is given by the SM particles plus $N_{\rm seesaw}$ seesaw RH neutrinos.
Notice that at least two heavy seesaw neutrinos are necessary to reproduce the solar and atmospheric neutrino mass scales so that $N_{\rm seesaw} \geq 2$. For definiteness we consider the case $N_{\rm seesaw} = 2$. 

With this splitting, the seesaw neutrinos will now not get their mass through the $\s$-phase transition. They could still get their mass via the $\eta$-phase transition though, in this case one could even imagine a double peaked GW spectrum, with one peak at high frequencies, from the $\eta$-phase transition, and one peak at low frequencies, from the $\s$-phase transition. 

Let us, more specifically, consider $T_\star \lesssim 100\,{\rm keV}$ since in this way the GW spectrum 
can be testable with Pulsar Time Arrays. 
We could so far consider the dark sector in equilibrium with the visible sector thanks to the Yukawa interactions.
However, now the Yukawa interactions of RH neutrinos undergoing the phase transition are not strong enough
to enforce equilibrium. One can still assume that the dark sector was coupled to the visible sector at temperatures
$T \gtrsim M$ and decoupled afterwards so that $T_{\rm dec} \sim M$.  Therefore, 
we have now to consider a dark sector phase transition temperature  $T'_\star \neq T_\star$.

Assuming entropy conservation, the ratio $r_T^\star \equiv T'_\star/T_\star$ can be simply calculated from
\be\label{rT}
r_T^\star=  \left[ {g^{\rm SM}_s(T_\star) \over g^{\rm SM}_s(T_{\rm dec})} \right]^{1/3} \,  .
\ee
For example, assuming $T_{\rm dec} \gtrsim 100\,{\rm GeV}$, one finds $r_T^\star \simeq 0.33$. This strongly decreases
the strength of the phase transition and even with the auxiliary scalar one gets GW spectra below 
the sensitivity of planned experiments and many orders of magnitudes below the putative NANOGrav signal. 
Some examples are shown in Fig.~3.
\begin{figure}
\begin{center}
        \psfig{file=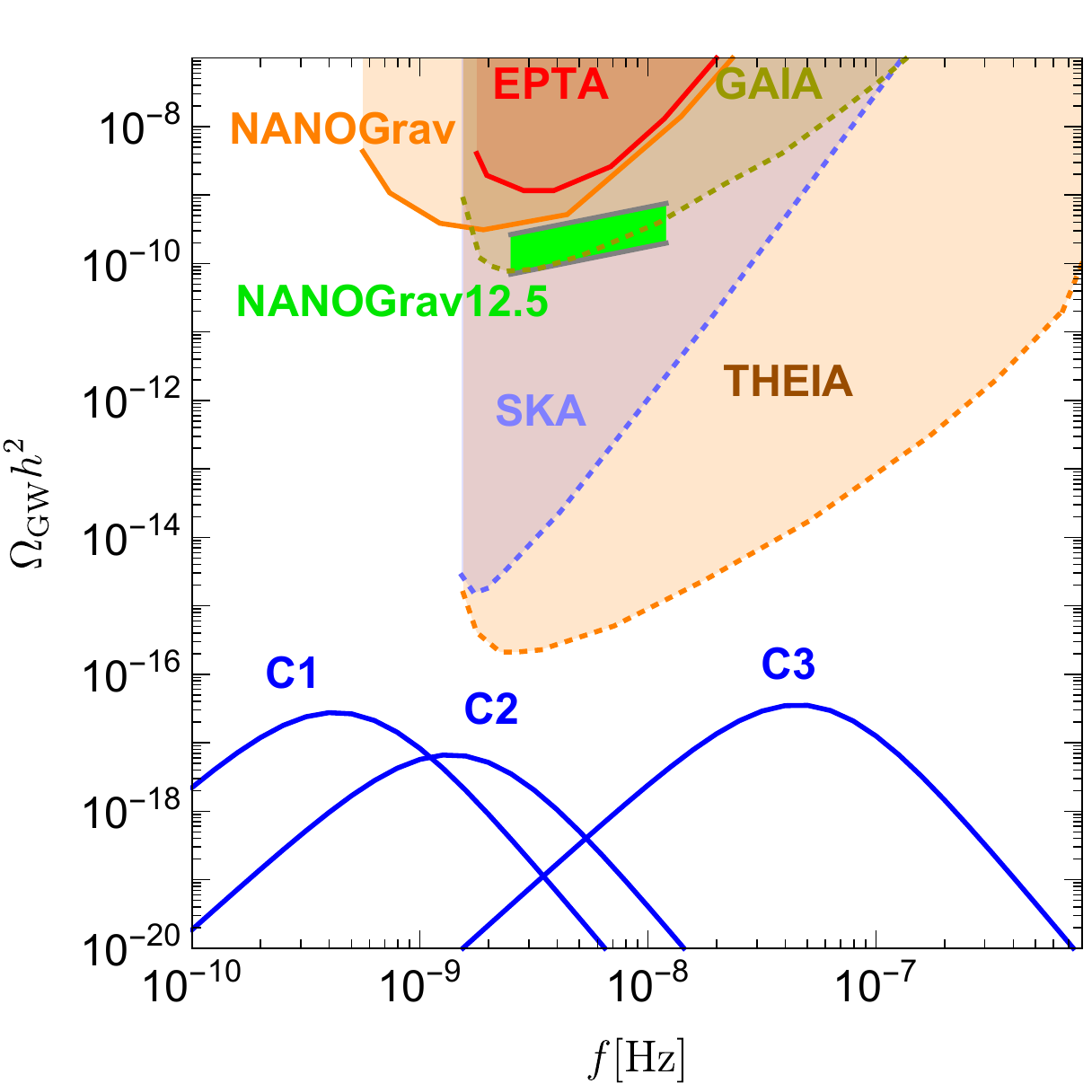,height=65mm,width=80mm}
\caption{Benchmark GW spectra in the low scale scenario for $T_{\rm dec} \gtrsim 100\,{\rm GeV}$
and $T_{\star} \lesssim 100\,{\rm keV}$ corresponding to $r_T^\star = 0.33$ (from \cite{DiBari:2021dri}).}
\label{fig:gws}
\end{center}
\end{figure}
If the seesaw scale, and consequently the decoupling temperature, is lowered to $\sim {\rm GeV}$, then
$r_T^\star$ increase to $0.4$ and the signal increases by a factor $\sim 30$ that is however not yet
sufficient to produce GW spectra within the sensitivity of planned experiments. However, this shows that
the signal is very sensitive to the value of $r_T^\star$ and, therefore, one could think to consider simply
the case $r_T^\star =1$. In this case one can imagine some process able to yield 
{\rm rethermalisation}  of the dark sector. 

On the other hand , there are cosmological constraints on the dark radiation amount to be considered that place
an upper bound on $r_T^\star$. These can be expressed in terms of the effective number of neutrinos species
$\D N_\nu^{\rm eff}(T)$ defined by
\be
g_{\rho}(T) = g^{\rm SM}_{\rho}(T) + {7\over 4}\,\D N_\nu^{\rm eff}(T)\,\left({T_\nu \over T}\right)^4 \, .
\ee
The asymptotic value of the SM contribution reached when temperature drops
below the electron mass is given by  $g^{\rm SM}_{\rho 0} \simeq 3.36$.
Cosmological observations place an upper bound on $\D N_\nu^{\rm eff}(T)$ at different temperatures:
\begin{itemize}
\item From $Y_{\rm p}$+CMB $\eta_B$ measurements $\Rightarrow \; \D N_\nu^{\rm eff}(t_{\rm f}\sim 1\,{\rm s}) \leq 0.5 \;(95\%{\rm C.L.})$;
\item From $D/H$+CMB $\eta_B$ measurements $\Rightarrow \; \D N_\nu^{\rm eff}(t_{\rm nuc}\sim 300\,{\rm s}) \leq 0.4 \;(95\%{\rm C.L.})$;
\item From CMB temperature+polarization anisotropies $\Rightarrow \; \D N_\nu^{\rm eff}(t_{\rm rec}) \leq 0.3 \;(95\%{\rm C.L.})$.
\end{itemize}
We have now to calculate $\D N_{\nu}^{\rm eff}(T)$ in the model and impose these upper bounds.
If the dark sector decouples at $T_{\rm dec}$ and does not recouple afterwards, then one has
\be
\D N_\nu^{\rm eff} = {4 \over 7}\,\left({11\over 4}\right)^{4\over 3}  \, r_T^4 \,g_\rho^{\rm dark sector}(T) = {\rm const}\,  ,
\ee
where $g_\rho^{\rm dark sector}(T) = g^{\eta+J+S}+ 7\,N'/4 = 3 +7N'/4$. Since it is constant throughout 
BBN and recombination, then all bounds apply and of course the most stringent is the one coming from CMB temperature anisotropies at recombination. These are the results for some values of $r_T^\star$ and $N'=1$:
\begin{itemize}
\item $r_T^\star = 0.33 \Rightarrow \D N_\nu^{\rm eff} \simeq 0.12$;
\item $r_T^\star = 0.4 \Rightarrow \D N_\nu^{\rm eff} \simeq 0.27$;
\item $r_T^\star = 0.6 \Rightarrow \D N_\nu^{\rm eff} = 1$;
\item $r_T^\star = 1 \Rightarrow \D N_\nu^{\rm eff} = 10$.
\end{itemize}
This shows that, very conservatively, we can at most take $r_T^\star = 0.6$. 
In Fig.~4 some benchmark GW spectra for this low scale scenario are shown. 
\begin{figure}
\begin{center}
        \psfig{file=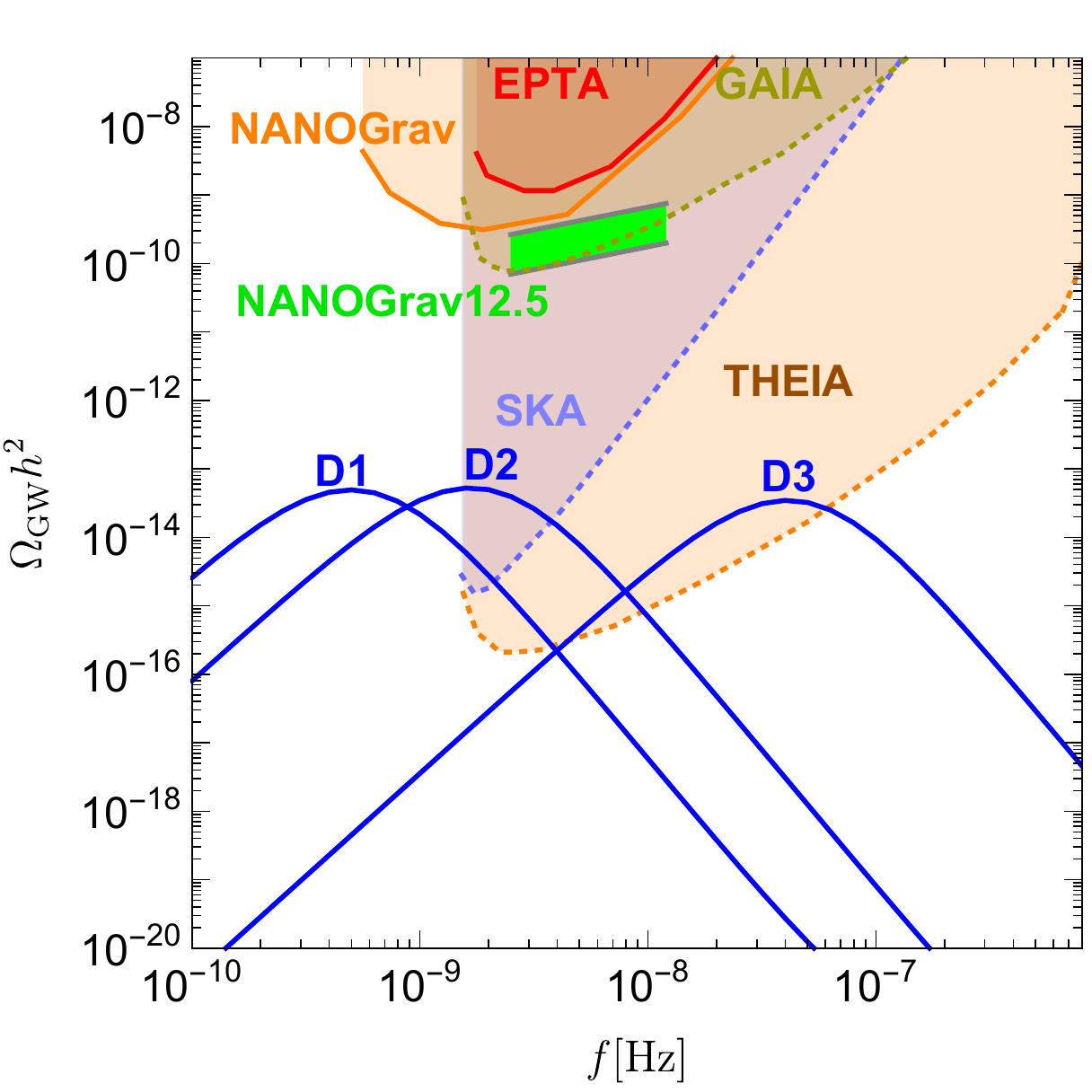,height=65mm,width=80mm}
\caption{Benchmark GW spectra in the low scale scenario for $T_{\rm dec} \gtrsim 100\,{\rm GeV}$
and $r_T^\star = 0.6$ (from \cite{DiBari:2021dri}).}
\label{fig:gws}
\end{center}
\end{figure}
One can see how in this case one can obtain signals detectable at planned experiments such as SKA  and THEIA
but still many orders of magnitude below the claimed NANOGrav signal. 

This low scale scenario with $r_T^\star =0.6$  is only very marginally allowed by cosmological constraints and moreover there is a question: how can one justify a partial rethermalisation process
leading to $r_T^\star = 0.6$? There is a more realistic scenario that can address both these issues and that moreover is also able to address a completely independent problem: the so-called {\em Hubble tension}. 
%In this way  such low scale phase transitions get motivated by different phenomenological problems. 

\section{GW signal and Hubble tension}

From CMB temperature and polarization anisotropies observations the {\em Planck} collaboration
derives the following value for the Hubble constant assuming the $\L$CDM model \cite{Planck:2018nkj}:
\be
H_0 = (67.66 \pm 0.42) \, {\rm km}\, {\rm s}^{-1} \, {\rm Mpc}^{-1} \,  .
\ee
On the other hand, astrophysical measurements of SN type Ia redshifts versus luminosity distances find \cite{Riess:2021jrx}:
\be
H_0 = (73.30 \pm 1.04) \, {\rm km}\, {\rm s}^{-1} \, {\rm Mpc}^{-1} \,  .
\ee
This $5 \sigma$ tension is the so-called {\em Hubble tension} \cite{Bernal:2016gxb}.

It was initially proposed that the injection of extra radiation prior to recombination at the
level of $\D N_\nu^{\rm eff} \simeq 0.5$ might provide a simple solution to solve the tension \cite{DiBari:2013dna}.
However, with more precise data it has now become clear that actually such an injection spoils the agreement with other
quantities such as $\sigma_8$ and in the end the global fit even worsens compared to the $\L$CDM model.
Many different solutions have been proposed and currently it seems that a class of solutions based on a modification of pre-recombination physics is more favoured \cite{Knox:2019rjx}.  The trick is to be able to reduce the sound horizon at
recombination without altering the well fitted CMB observables (i.e., the positions of the peaks). 
If we introduce an interaction  between the Majoron background and ordinary neutrinos (this is 
a modification of the early proposal made in \cite{Chacko:2003dt})  of the kind
\be\label{nudark}
-{\cal L}_{\rm \nu-{\rm dark}} = {i\over 2}\,\sum_{i=2,3}\lambda_i \, \overline{\nu_{i}} \, \g^5 \,\nu_{i} \, \eta + 
{i\over 2}\,\lambda_1 \overline{\nu_{1}}\,\g^5 \,\nu_{1} \, J  + {\rm h.c.}\,  ,
\ee
it has been shown that a fit of cosmological observations, including Hubble measurements, improves
compared to the $\L$CDM, though some tensions remains \cite{Escudero:2019gvw,Escudero:2021rfi}. 
With this interaction, dark sector and ordinary neutrinos equilibrate after the 
$\sigma$-phase transition so that $T' = T_\nu \simeq 0.6\,T$ and one has
$\D N_\nu^{\rm eff}(t_{\rm rec}) \simeq 0.5$ but 
$\D N_\nu^{\rm eff}(t_{\rm f}) \simeq \D N_\nu^{\rm eff}(t_{\rm nuc}) \simeq 0.1$. 
The GW signal can be moreover further enhanced increasing $N'$, though, as we discussed, there
is an upper limit $N' \lesssim 10$ from requiring the stability of the potential. 

In Fig.~5  some benchmark GW spectra for this low scale scenario are shown. 
\begin{figure}
\begin{center}
\psfig{file=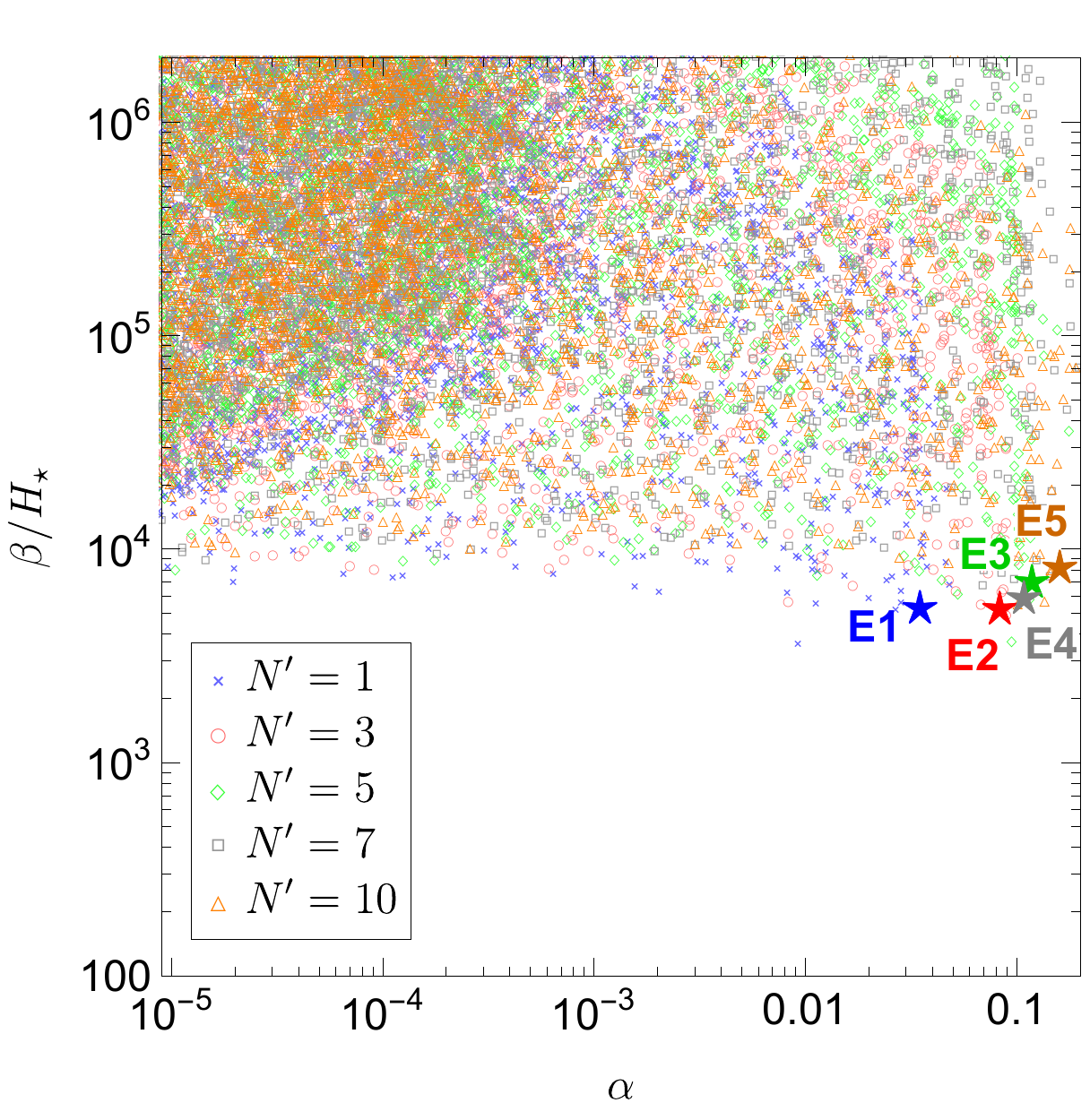,height=70mm,width=70mm} 
\hspace{2mm}
\psfig{file=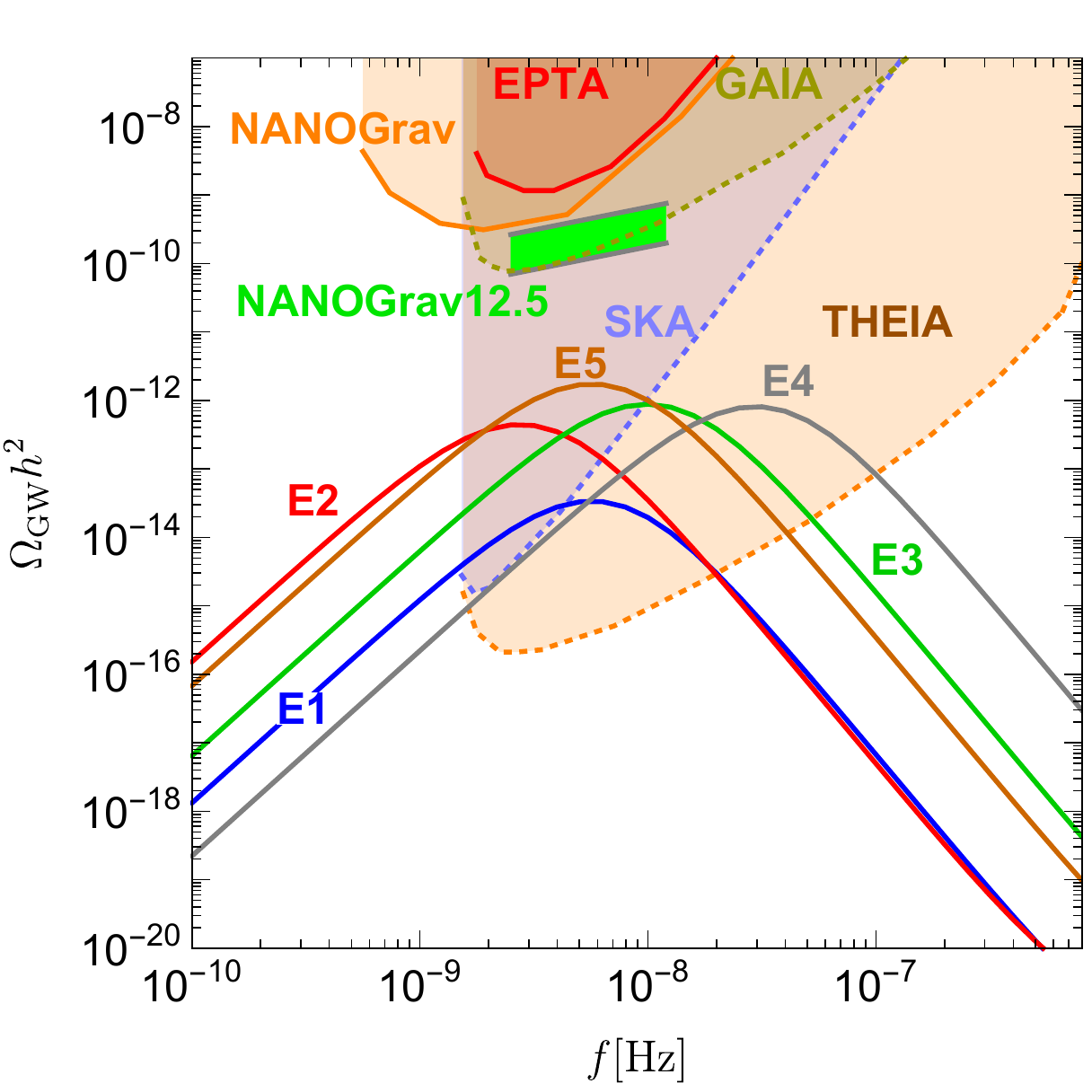,height=70mm,width=70mm} 
\end{center}
\caption{Rethermalization scenario addressing the Hubble tension.  Left panel: Scatter plot in the $(\a, \beta/H_\star)$ plane.  
 Right panel: GW spectra for the five benchmark points E1--E5 in the left panel corresponding to $N' =1 ,3, 5, 7, 10$, respectively, and marked with stars (from \cite{DiBari:2021dri}).}
\label{f6}
\end{figure}
One can see how there can be a significant enhancement by increasing $N'$ since in this way
the relative weight of the dark sector compared to the visible sector gets higher and accordingly the
strength of the phase transition $\a$. However, even for a maximum value $N'=10$ the peak of the GW spectrum is 
about two orders of magnitude below the NANOGrav signal. 

\section{Dark matter from a SFOPT}

In addition to gravitational waves, it is also possible to conceive production of dark matter during a SFOPT 
generating a Majorana mass term \cite{DiBari:2020bvn}. If the following 5-dimensional Anisimov  operator \cite{anisimov,ad}      
\be
{\cal L}_{\rm DS} = {\lambda^{\rm mix}_{\rm DS}\over \Lambda_{DS}}\,\Phi^\dagger \, \Phi \, \overline{N_{\rm D}^c} \,{N_{\rm S}}
\ee
is added to the Lagrangian in Eq.~(\ref{eq:L_l}), it can induce a mixing between a seesaw neutrino $N_{\rm S}$, acting as
a {\em source} RH neutrino, and a heavy neutral lepton in the dark sector $N_{\rm D}$, that can be regarded as
a {\em dark}  RH neutrino. During the phase transition the source RH neutrino mass varies growing from zero to its final value. If the phase transition is first order and bubbles nucleate, then at the bubble wall a resonance can be met
and this will trigger a production of dark matter, in addition to GWs. In order for the production to be efficient enough
to generate an abundance that can reproduce the observed dark matter abundance, the coupling $\lambda^{\rm mix}_{\rm DS}$ 
has to be sufficiently large but this also makes $N_{\rm D}$ too short lived to satisfy CMB temperature anisotropy constraints, imposing a lower bound $\tau_{\rm DM} \gtrsim 10^{25}\,{\rm s}$ for masses below the $200\,{\rm GeV}$ (at higher masses neutrino telescope lower bound is even more stringent, $\tau_{\rm DM} \gtrsim 10^{28}\,{\rm s}$).

For this reason $N_{\rm D}$ cannot be the dark matter but it can decay into a lighter dark RH neutrino 
via a second Anisimov operator inducing a mixing between $N_{\rm D}$ and $N_{\rm DM}$, explicitly
\be
{\cal L}_{\rm D-DM} = {\lambda^{\rm mix}_{\rm D-DM}\over \Lambda_{D-DM}}\,\Phi^\dagger \, \Phi \, \overline{N_{\rm D}^c} \,{N_{\rm DM}} \,  .
\ee
This lighter RH neutrino can now play the role of dark matter and satisfy observational constraints.
In this way one can have at the phase transition a simultaneous genesis of Majorana masses, GW stochastic background and 
dark matter. 

\section{Summary}

Origin of (Majorana) neutrino masses in a SFOPT within a Majoron model can not only reproduce neutrino masses with type-I seesaw models and address cosmological  origin of matter (the dark sector can also provide the dark matter while seesaw neutrinos can produce the matter-antimatter asymmetry with leptogenesis) but it might also give rise to a  GW stochastic background. We have discussed scenarios where the spectrum can be within the sensitivity of future GW interferometers
and in this respect a second auxiliary scalar seems to be a crucial ingredient, as already noticed in the case of EWPT. 

Low scale scenarios testable with PTAs are also motivated by the existence of tensions within the LCDM model, such as the Hubble tension. Though NANOGraV signal seems out of reach, theoretical uncertainties are still very large 
\cite{Croon:2020cgk,Guo:2021qcq} and all results should be regarded currently as indicative (in particular, see \cite{Jinno:2021ury}  as a possible way to enhance the signal taking into account primordial fluctuations.)

Finally, I wish to highlight that in traditional high scale seesaw scenarios with Majorana masses that can be as high as 
the grand-unified scale, one would expect GW spectra peaking in the ultra high-frequency range (MHz-GHz). For this reason
the models we present certainly strongly motivate exploring  new concept of detectors able to cover this range 
\cite{Aggarwal:2020olq}.

\newpage
%\vspace{-4mm}
\subsection*{Acknowledgments}

I wish to thank Danny Marfatia and  Ye-Ling Zhou  for a fruitful collaboration;  Graham White for 
useful comments and to draw my attention to relevant references; Francesco Muia and Fernando Quevedo for
useful discussion on the opportunity to develop very high frequency detectors to test phase transitions at very high scales.
I acknowledge financial support from the STFC Consolidated Grant ST/T000775/1.

\end{document}